# Application of the Phase Correction Method in Tracking Modules of Navigation Receivers

Kudryavtsev I.A., Shafran S.V., Balmasova M.A.
Samara National Research University

Abstract

Tracking modules are very important parts of GNSS receivers, providing precise monitoring of satellite signals and evaluating time delays and frequency shifts necessary to calculate ultimate position. Common architecture contains two tracking loops: PLL and DLL, generating local copies of the PRN code and carrier frequency in order to ensure close correlation with the received signal. One of the problems of these loops is their significant inertia in case of narrow bandwidth, which is desirable to mitigate signal noise.

The paper describes a method of phase correction, improving accuracy and dynamic characteristics of the tracking process in GNSS receivers. The pivotal idea of the method supposes correction of the phase deviation between the received signal and its local copy at the end of every integration period. This correction is performed directly over the phase accumulator of a DDS system. This approach reduces inertial effects crucially.

The results of numerical simulations and physical experiments display the efficiency of the method in the case of tracking using long-time integration. Using deviation values obtained from unfiltered input signals is a drawback of the approach, so in the case of relatively short integration intervals, such correction can provide worse results; however, using narrower filter bandwidth and longer timing, which are available due to less inertia, outweighs this drawback.

Using narrow filter bandwidth requires attention to transient processes, described in the paper. It is recommended to start tracking with the integration intervals of a few milliseconds and proceed to longer ones after some time, when the PLL has stabilized and the signal frequencies are not fluctuating significantly. This process is similar in PLL and DLL; however, DLL filter bandwidth has more influence on the ultimate accuracy, based on code measurements.

Application of phase correction method with narrow filter bandwidth is much more efficient when FLL-assisted PLL tracking is used. It is shown that in this case a separate FLL discriminator is not needed, and the resulting filter structure can be significantly simplified.

In conclusion results of experiments are discussed, including influence of filter bandwidth on inertial characteristics of PLL and phase measurements. All the experiments were carried out with a second-order filters, however the method can be used with a third-order filter architecture as well.

## Introduction

Tracking modules are among the core elements of real-time Global Navigation Satellite System (GNSS) receivers. They are responsible for determining the instantaneous time delays of the signal propagation from navigation satellites to the receiver, extracting the navigation data transmitted by the satellites, and performing a number of auxiliary functions. Positioning accuracy directly depends on the accuracy of time delay estimation; therefore, considering the significant level of interference and noise affecting the navigation signal, one of the main means of improving accuracy is increasing the integration interval duration [1].

In the tracking modules of navigation receivers of GPS L1 and GLONASS L1/L2, the integration interval is limited by the navigation bit length. Furthermore, increasing the integration time increases the inertia of the carrier frequency and phase tracking loop, which can affect the dynamic characteristics of the receiver and increase its sensitivity to the time instability of the receiver's reference clock oscillator. In this regard, finding ways to reduce this inertia in the carrier tracking loop seems highly relevant.

## Problem Statement

The architecture, operating principles, and characteristics of tracking modules are discussed in a number of publications, such as [1-3, 6-11], which consider various aspects, including tracking accuracy, impact on positioning errors, etc. A simplified block diagram of a tracking module processing signals from the RF Front-End output is shown in Fig. 1. It implements continuous monitoring of the received signal by computing the correlation integral over a selected integration interval. Then, using a discriminator (D), the frequency and/or phase mismatch is determined, which, after being processed by a loop low-pass filter (LPF), is fed to the local carrier Numerically Controlled Oscillator (NCO). The most common implementation of the tracking loop is based on the Phase-Locked Loop (PLL), which includes a phase discriminator, although in some cases, a Frequency-Locked Loop (FLL) or a combined method is used. This paper is limited to the PLL (and FLL-assisted PLL) implementation, which strives to minimize the phase mismatch between the received and local signals. Simultaneously, in the same manner, the local ranging code generator is adjusted in the Delay-Locked Loop (DLL), which in most receivers plays the main role in calculating the time delay. However, the phase mismatch between the carrier of the received and local signals can also cause an increase in tracking errors in the DLL.

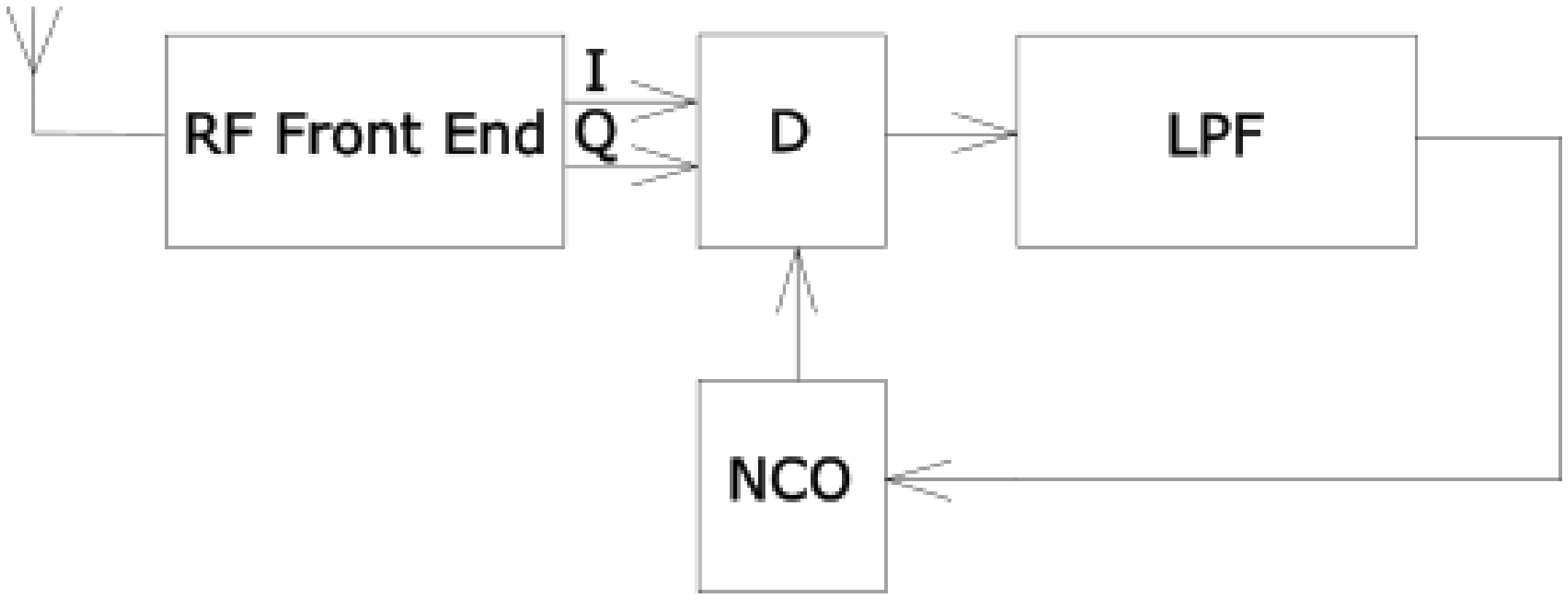

**Fig. 1.** Simplified structure of a tracking module

The tracking algorithm includes a crucial stage where the tracking module must acquire the signal using the Doppler shift and code delay estimates obtained from the signal acquisition system. Here, two factors are important: first, the acquisition system forms approximate estimates of these parameters; second, when the local carrier oscillator is started, the phase shift between its signal and the received signal can have a random value, in particular due to the delay caused by processing in the acquisition module. Fig. 2 shows an example of a transient process in the tracking loop, where an overshoot is observed, associated with the random value of the phase difference between the received signal and the local carrier oscillator signal at the beginning of the process. This result was obtained through simulation, where, for illustrative purposes, the initial frequency of the local oscillator and the received signal were set to be equal, but the initial phases differed, while the input signal frequency remained constant. The PLL loop filter bandwidth was set to 25 Hz. The signal with an artificially high Carrier-to-Noise density ratio ${C}/{N_0} = 70\,dbHz$ is shown for comparison with a signal having a ${C}/{N_0} = 45\,dbHz$ value, which is close to the normal level of the GPS signal. The simulation used a real tracking module built on a classic architecture in the Vivado system.

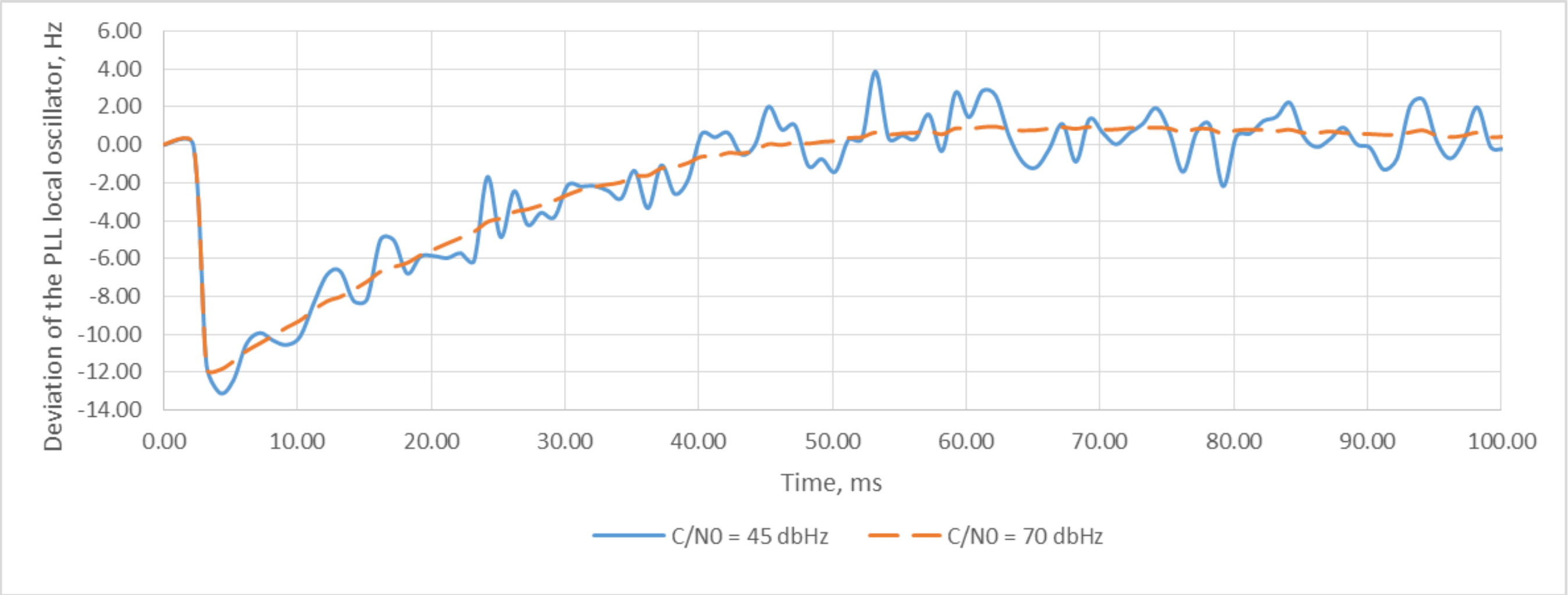

**Fig. 2.** Transient process in the PLL with zero initial frequency deviation

At the same time, in a PLL implementation, the range of allowable deviations is limited by the discriminator's linear range. The issues of discriminator selection are detailed, for example, in [2]. In a Costas loop, which is insensitive to phase transitions, the phase difference is limited to the range of $\left[-\frac{\pi}{2}, \frac{\pi}{2}\right]$.

Thus, tracking in the PLL is stable when the phase differences do not exceed these boundaries, which also limit the frequency difference, considering the integration interval duration. The obvious limitation for the Costas loop can be described by the expression: $|\Delta\omega| \cdot T < \frac{\pi}{2}$, where $\Delta\omega$ is the frequency difference (rad/s), and t is the integration interval (s).

In addition, the desire to set the filter noise bandwidth as narrow as possible in order to reduce tracking errors leads to the inertia of the circuit as a whole, which in turn can lead to a loss of lock, even if the above limitation is met, due to the slow compensation of the phase mismatch, which leads to the accumulation of error. The PLL generally strives to minimize the phase mismatch; however, due to inertia, the phase difference can reach the limiting value of the discriminator, causing the output signal of the discriminator to change polarity. In such a case, the loop can enter the mode of relaxation oscillations. The issues of PLL stability are detailed in [2, 3]. An example of such a transient process is shown in Fig. 3. A filter with a bandwidth of 50 Hz was used for the simulation.

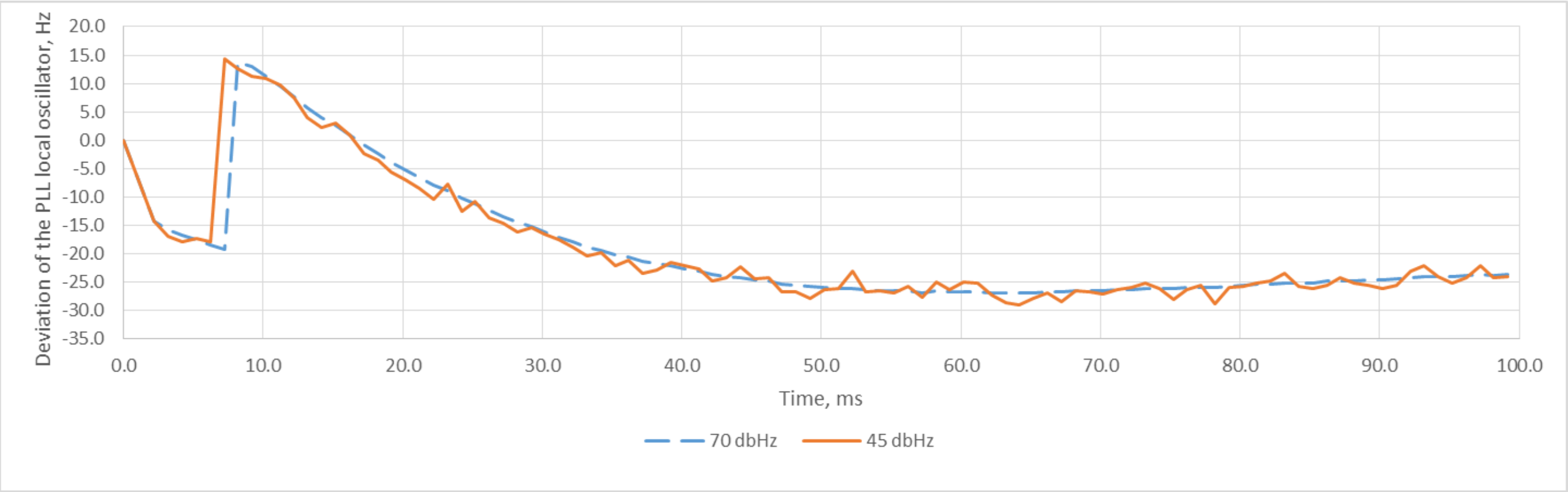

**Fig. 3.** Transient process in the PLL with nonzero (22 Hz) initial frequency deviation

It can be seen from Fig. 3 that there is a frequency jump caused by the above-mentioned effect in the phase discriminator. With too large a frequency, the module will not be able to acquire the signal. In general, the transient process can depend on the SNR and the initial phase difference between the received signal and the local oscillator. The frequency deviation of the local oscillator shown on the ordinate axis represents the deviation from the initial value obtained from the acquisition system.

Fig. 4 shows an example of signal acquisition by the module with an initial frequency deviation of 70 Hz.

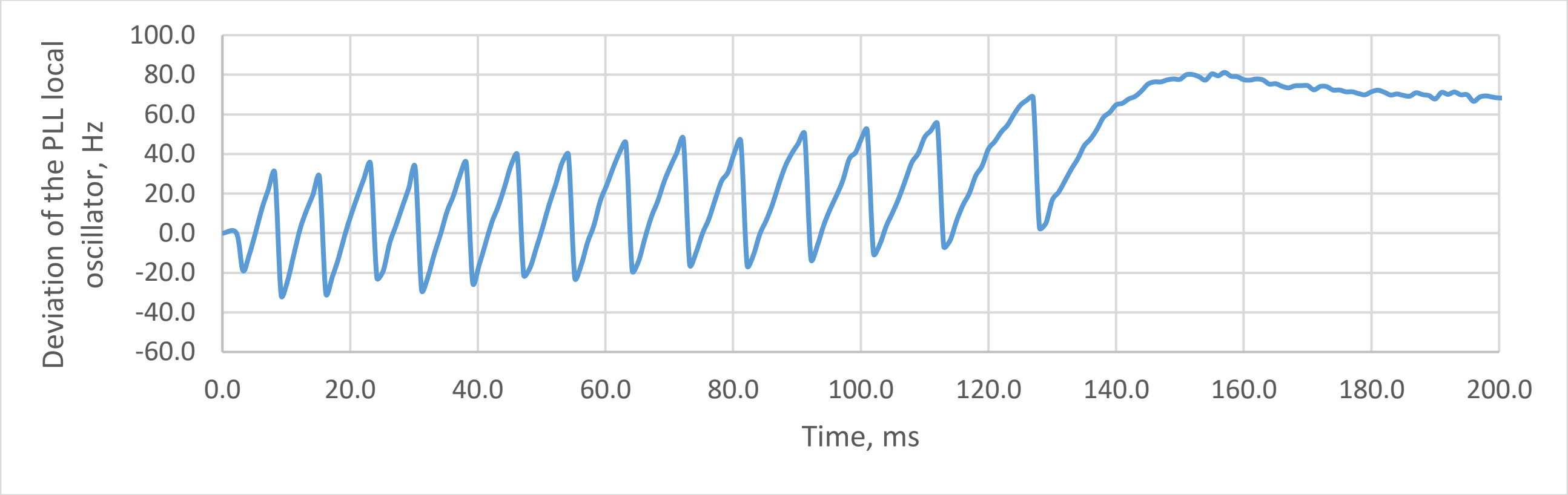

**Fig. 4.** Transient process with relaxation oscillations in the PLL with initial frequency deviation of 70 Hz

Thus, the modification of the PLL, where the phase error is not accumulated, could be a possible solution. It is worth noting that the issues concerning the development of frequency synthesizers are discussed in various papers, not necessarily focused on GNSS systems (see, for example, [12]).

### Phase Shift Compensation Method

When implementing the local carrier oscillator based on Direct Digital Synthesis (DDS) technology [4, 5], frequency control reduces to modifying the phase accumulator increment. The frequency change is determined by the expression: $\Delta f_{LOC} = \frac{M}{2^n} \cdot F_0$, where M – is the increment value, n is the accumulator bit width, and $F_0$ is the reference oscillator frequency. The simplified block diagram of such a generator is presented in Fig. 5. The issues of selecting DDS-based generator parameters are detailed, for example, in [4].

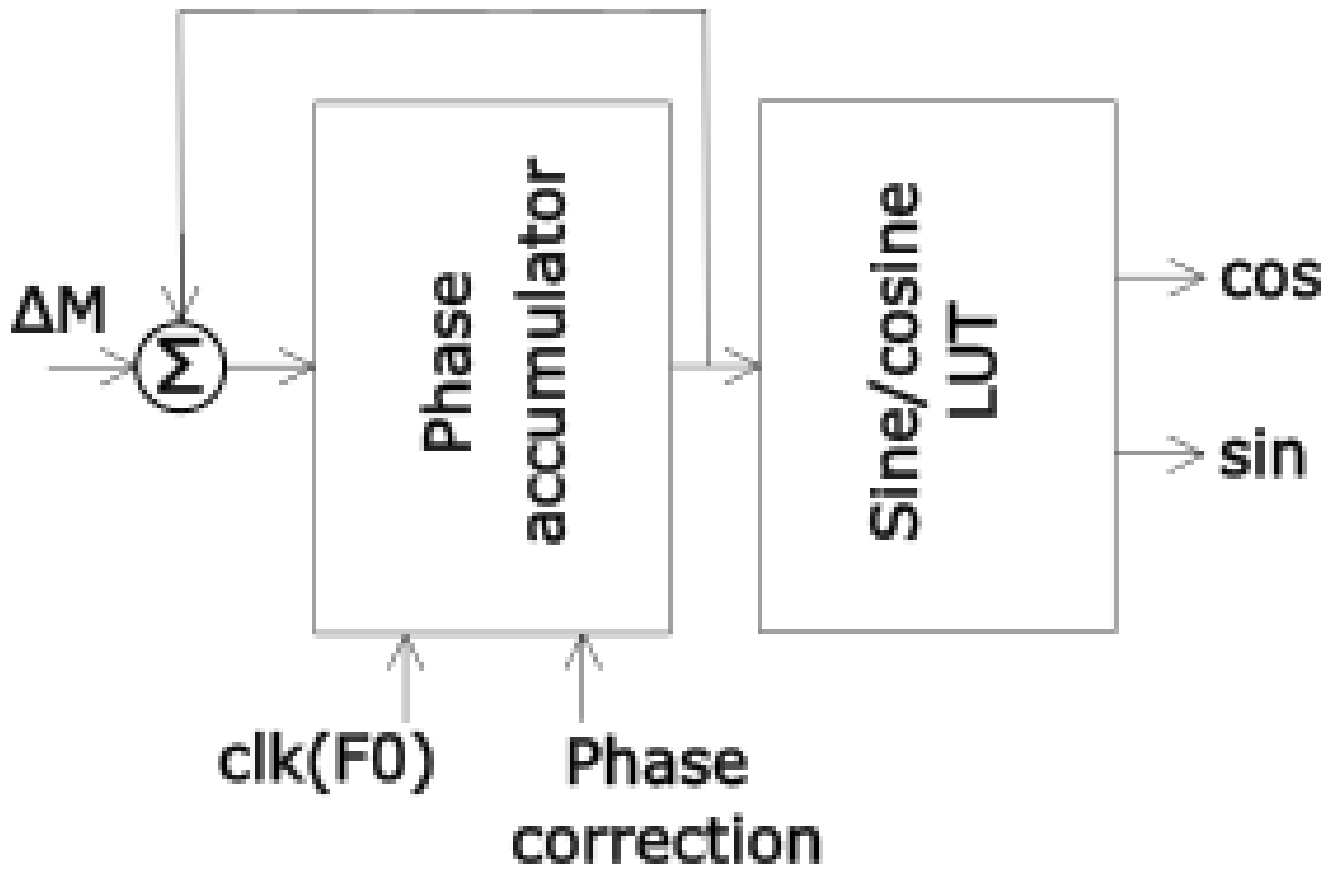

**Fig. 5.** Simplified block diagram of the local carrier oscillator in the PLL

An important aspect is the frequency of the accumulator increment update, which is determined, in turn, by the integration interval in the tracking module. For example, for a GPS L1 receiver, when the integration interval is 20 ms, the increment is updated 50 times per second (at the beginning of each interval), after which the DDS frequency remains unchanged until the end of the integration.

To compensate the phase shift directly, without changing the accumulator increment value, it is necessary to modify the content of the phase accumulator itself. For this purpose, a separate correction input is provided, as shown in Fig. 5. The correction value is added directly to the phase accumulator.

The phase shift value $\Delta\varphi$ is calculated using the integration results in the in-phase and quadrature channels (IP и QP). Their values are determined as follows:

$$IP + jQP = \int_0^T e^{j(\omega_{in}t+\Delta\varphi)} \cdot e^{-j\omega_{loc}t} dt$$

When the input signal frequency $\omega_{in}$ is equal to the local oscillator frequency $\omega_{loc}$ we get:

$$IP + jQP = \int_0^T e^{j\Delta\varphi}\, dt = [\cos(\Delta\varphi) + j sin(\Delta\varphi)]\, T$$

From the expression above we can derive $\Delta\varphi = \text{atan}(\frac{QP}{IP})$. To convert this to the EPLL (the signal at the filter input), it is necessary to multiple $\Delta\varphi$ by a normalizing factor $\frac{1}{2\pi}$.

Changing the accumulator increment by ΔM is equivalent to shifting the carrier frequency by $\Delta f = \frac{\Delta M}{2^n} \cdot F_0$, where $F_0$ is the clock frequency. During the integration interval $T$ the total change in the phase accumulator will be $\Delta M \cdot F_0 \cdot T = \frac{\Delta f \cdot 2^n}{F_0} \cdot F_0 \cdot T = EPLL \cdot 2^n$ that corresponds to a phase increment of $EPLL = \Delta f \cdot T$.

Thus, the phase correction of the carrier generator reduces to correcting the phase accumulator by the value of $EPLL \cdot 2^n$.

To compensate for the initial phase difference, it is advisable to implement phase correction at the beginning of tracking. Fig. 6 shows the change in the transient process with such correction. The same signal was used as in the case shown in Fig. 2. It can be noted that the overshoot present in Fig. 2 is absent here.

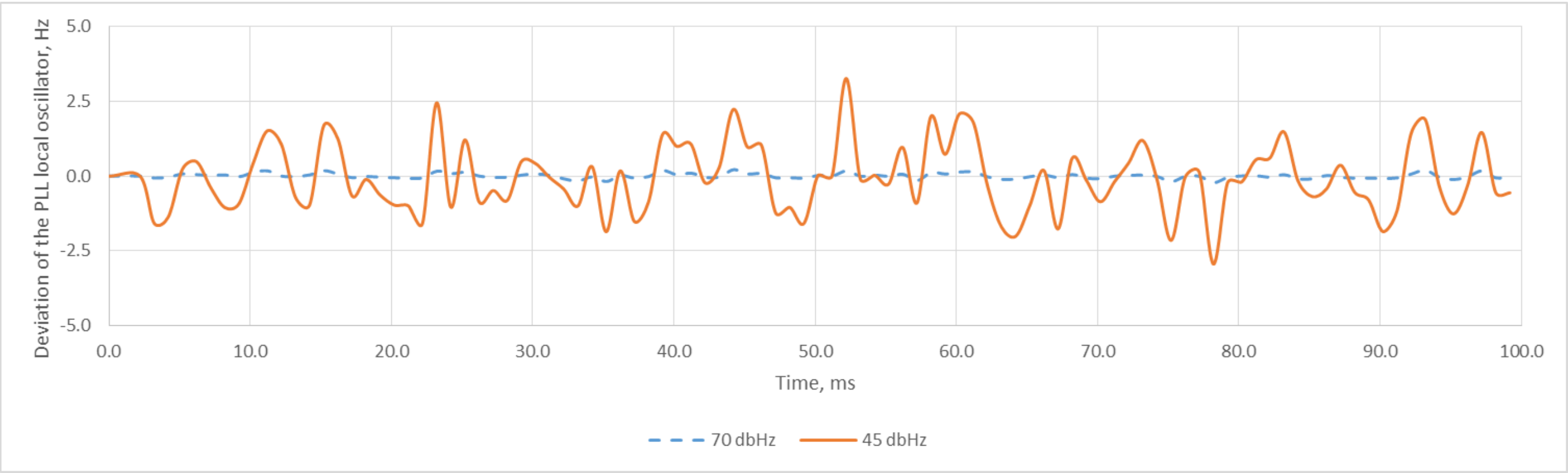

**Fig. 6.** Transient process with initial phase correction and zero frequency deviation

In the presence of a frequency offset (let $\omega_{in} = \omega_{loc} + \Delta\omega$) we can write:

$$IP + jQP = \int_0^T e^{j(\omega_{loc}t + \Delta\omega t + \varphi_{in})} \cdot e^{-j\omega_{loc}t} dt = \int_0^T e^{j\Delta\omega t} \cdot e^{j\varphi_{in}} dt$$

Hence, we obtain:

$$P + jQP = e^{j\varphi_{in}} \int_0^T e^{j\Delta\omega t} dt = e^{j\varphi_{in}} \int_0^T (cos(\Delta\omega t) + jsin(\Delta\omega t)) dt = \frac{e^{j\varphi_{in}}}{\Delta\omega} \cdot [sin(\Delta\omega T) - j(cos(\Delta\omega T) - 1)]$$

$$IP + jQP = \frac{e^{j\varphi_{in}}}{\Delta\omega} \cdot \left[2 \sin\left(\frac{\Delta\omega T}{2}\right) \cos\left(\frac{\Delta\omega T}{2}\right) + j2sin^2\left(\frac{\Delta\omega T}{2}\right)\right] =$$

$$= \frac{2e^{j\varphi_{in}}}{\Delta\omega} \cdot \sin\left(\frac{\Delta\omega T}{2}\right) \cdot \left[cos\left(\frac{\Delta\omega T}{2}\right) + j \sin\left(\frac{\Delta\omega T}{2}\right)\right]$$

$$\frac{QP}{IP} = \frac{\sin\left(\varphi_{in} + \frac{\Delta\omega T}{2}\right)}{\cos\left(\varphi_{in} + \frac{\Delta\omega T}{2}\right)} = \tan\left(\varphi_{in} + \frac{\Delta\omega T}{2}\right)$$

It follows that the output signal of the phase discriminator in the Costas loop is proportional to the sum of the initial phase deviation and $\frac{\Delta f T}{2}$. In other words, the application of the phase correction described above can be interpreted as an additional frequency offset of the local oscillator. If a noise component $\Delta\varphi$ is present in the discriminator output signal, it will automatically be translated into an equivalent change in the local oscillator frequency, approximately equal to (in Hz): ${}^{\Delta\varphi}/_{\pi T}$, where $\Delta\varphi$ is given in radians.

Fig. 7 shows the result of the local carrier oscillator phase correction at the beginning of signal acquisition by the tracking module, obtained in the simulation mode of the digital model. In this case, for maximum illustrative purposes, the phase-modulated input signal was simulated with minimal noise, the starting frequency of the local oscillator was chosen with zero deviation from the input signal frequency, but the phase of the input signal had a random value. The correction was performed at the end of the first integration period immediately after receiving the signal from the PLL discriminator output.

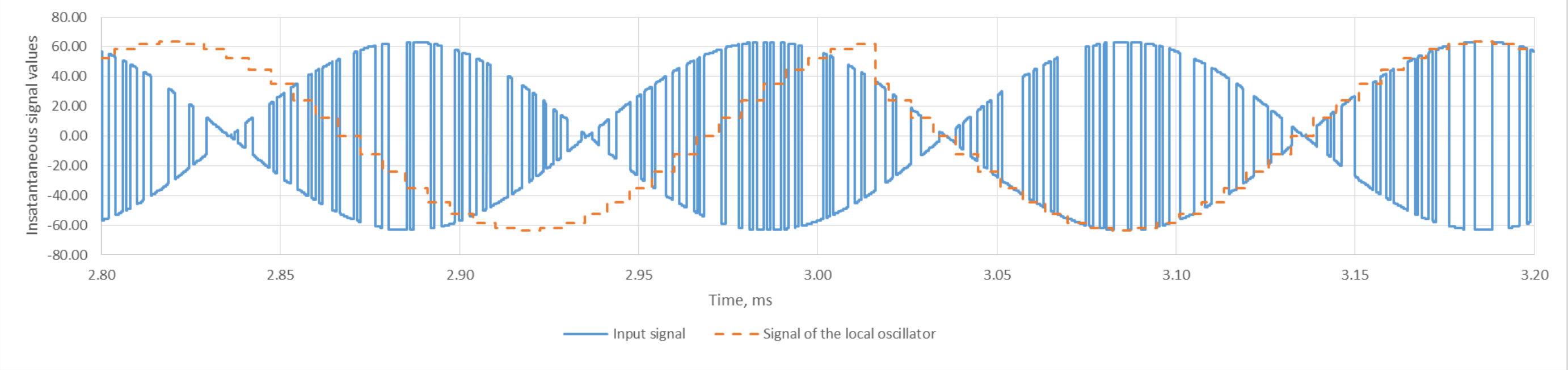

**Fig. 7.** Phase correction of a local carrier oscillator

Fig. 7 shows the moment of correction, after which the phase of the local generator signal is aligned with the phase of the input signal envelope. In the presence of noise and frequency deviations, the correction may be incomplete due to the error in determining the phase difference by the discriminator; however, practice shows the effectiveness of the method in this case as well.

When applying phase correction directly in every integration cycle, it is also necessary to modify the structure of the PLL loop filter, which must consider this correction; otherwise, the duration of the transient process will increase, which is explained by a decrease in the error value (phase difference) and, accordingly, the rate of local oscillator frequency correction.

The scheme of the filter used in this case is shown in Fig. 8. In fact, this is a combined second-order filter corresponding to the FLL-assisted PLL concept [2]. At the same time, the frequency error value is determined, as indicated above, without the need to use a separate discriminator. Moreover, considering the fact that the frequency error can be directly obtained from the phase discriminator signal, the filter can actually be reduced to a standard second-order PLL filter by modifying the coefficient $\omega_{0p}^2$.

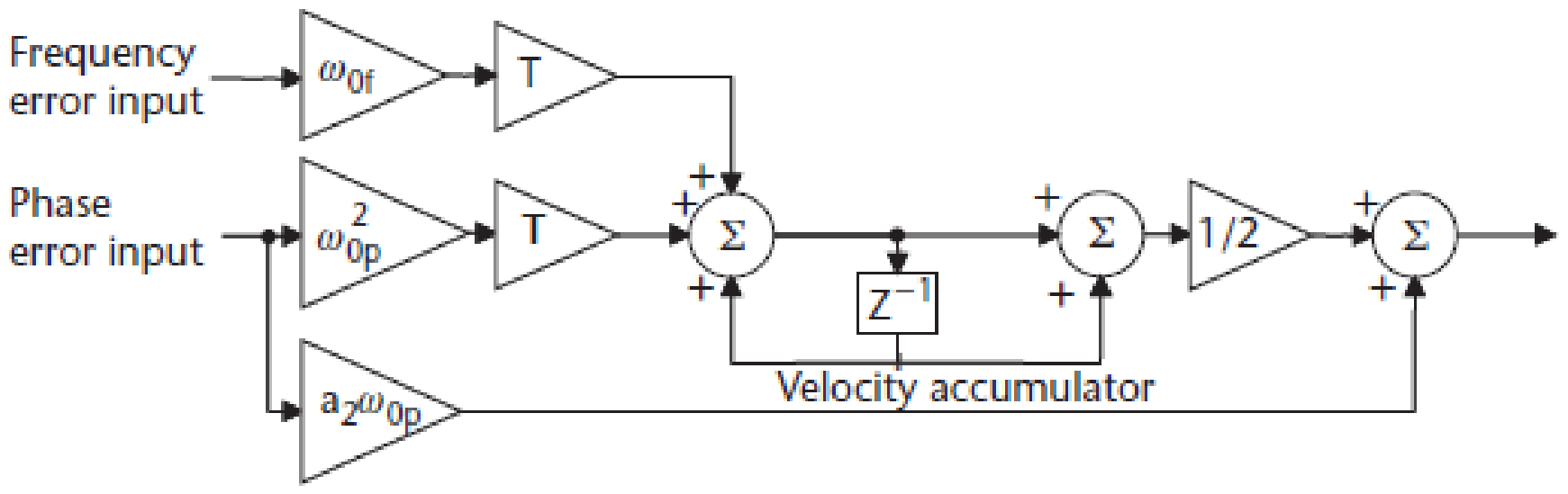

**Fig. 8.** Schematic diagram of a FLL-assisted PLL filter [2]

Since, according to [2], $\omega_{0f} = \frac{B_n}{0.25}$, a $\omega_{0p}^2 = \left(\frac{B_n}{0.53}\right)^2$, where $B_n$ is the filter bandwidth, $a_2$is the second–order coefficient (damping ratio), and also considering that the signal at the phase error input $EFLL = 2 \cdot \frac{EPLL}{T}$, we obtain that the filter presented in Fig. 8 is transformed to a standard second-order PLL filter [2], the scheme of which is shown in Fig. 9, taking into account the replacement of the coefficient $\omega_0^2 = 2 \cdot \frac{B_n}{0.25T} + \left(\frac{B_n}{0.53}\right)^2$.

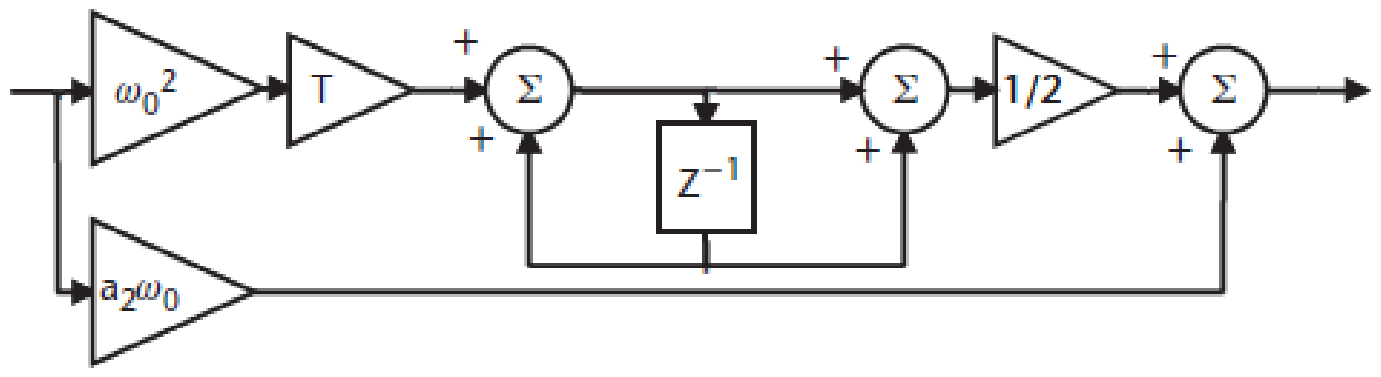

**Fig. 9.** Schematic diagram of a second-order PLL filter

Using such a combined filter (FLL-assisted PLL) improves tracking performance in general, as shown in [13]. Fig. 10 shows the graphs of the transient process when using phase correction with the above-described filter in the PLL loop.

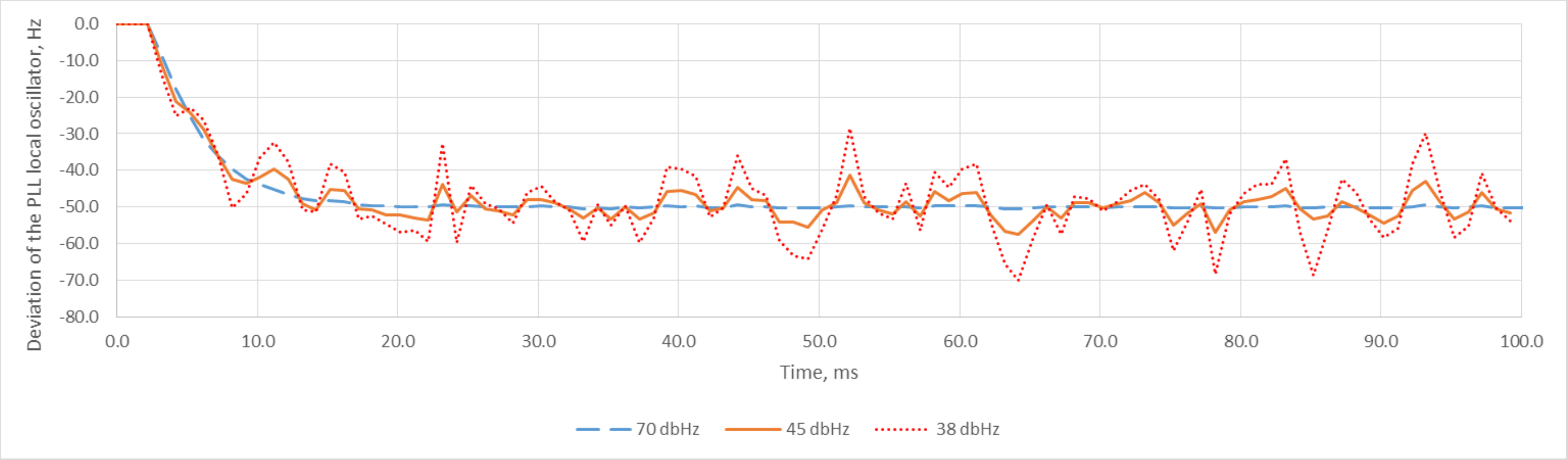

**Fig. 10.** Transient processes in the PLL with phase correction and zero initial phase shift

It should be noted that a tracking module with these improvements is significantly less sensitive to initial frequency deviation. Simulation was carried out for several initial frequency deviations, the results of which are shown in Fig. 11. In all cases, the filter noise bandwidth was set to 25 Hz with an integration time of 1 ms.

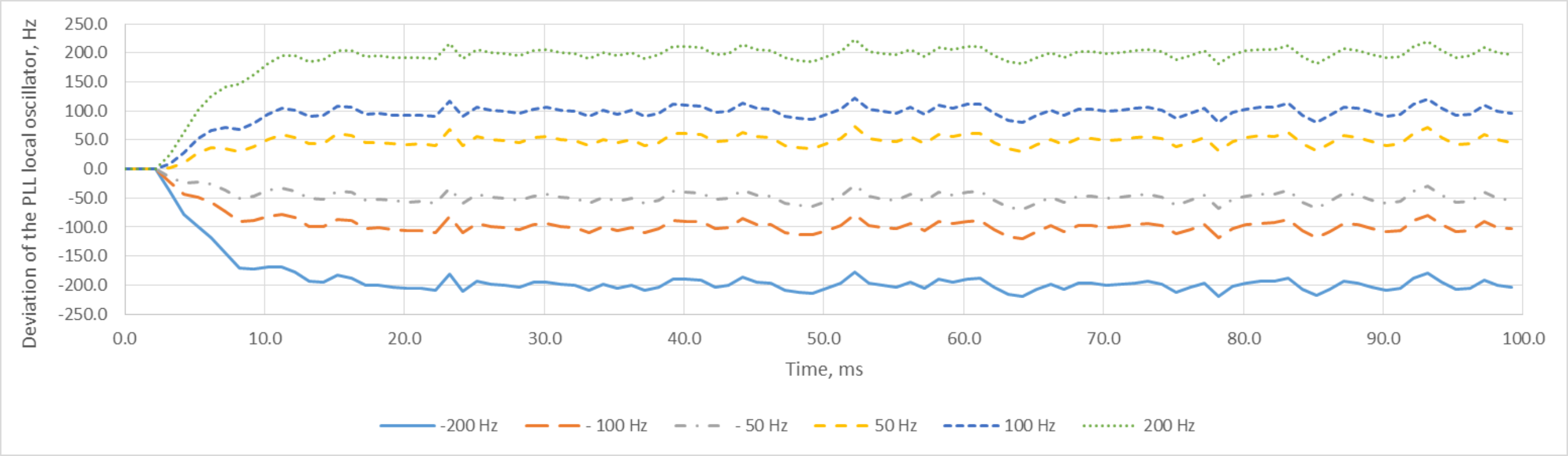

**Fig. 11.** Transient processes in the PLL with various frequency deviations

Fig. 12 shows the influence of the filter bandwidth on the transient process with initial frequency deviation of 100 Hz. In all cases, the same sample of signal samples was used with $C/N_0$ of approximately 38 dBHz.

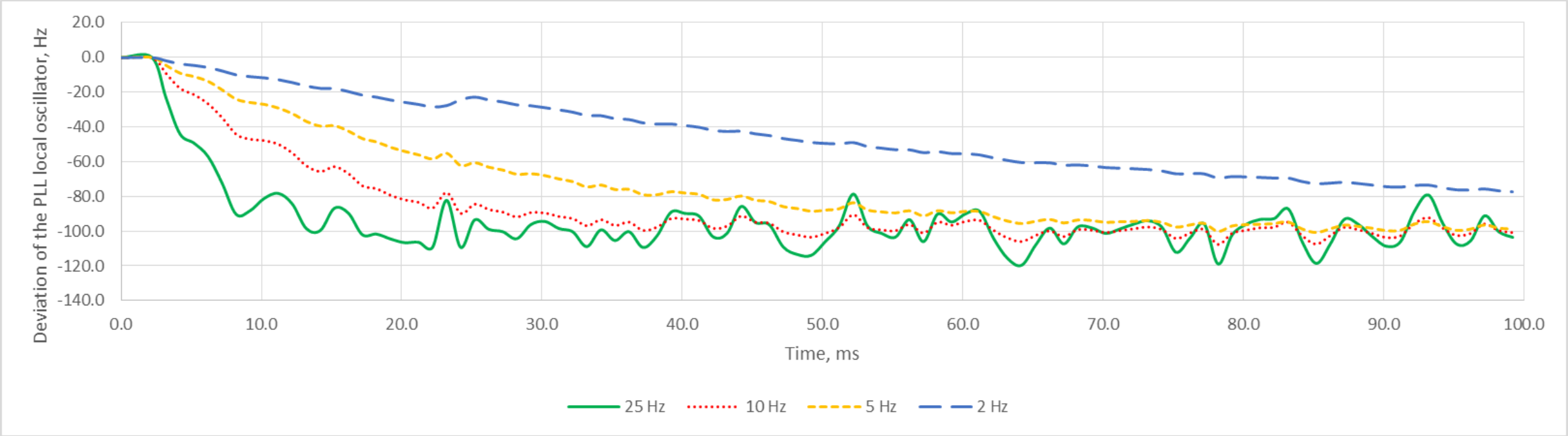

**Fig. 12.** Transient processes in the PLL with various filter settings

It should also be noted that the filter channel that does not have integrators obviously does not compensate phase deviations not associated with velocity, since the presence of correction automatically compensates them in every integration interval. Simulating a filter with a zero coefficient $a_2$ (the disabled branch of the circuit in Fig. 9, which does not have integrators) is difficult because the influence of this coefficient is noticeable only over significant intervals; therefore, the filter variant with a zero coefficient $a_2$ was investigated experimentally, as described below.

## Experimental Research

For experimental purposes, a navigation receiver implementing the above-described technology was developed, in which the simultaneous processing of the signal from the same navigation satellite (GPS L1 C/A) was performed by two tracking modules with different settings.

Fig. 13 shows the results of comparing the operation of the tracking module with the phase correction algorithm with the operation of a classic-type module. The filter noise bandwidth in both cases was set to 25 Hz, and the integration time was 1 ms. It can be seen that in this case, the phase correction expectedly introduces additional noise. The intensity of the noise depends on the SNR; the quantitative effect of this noise can be estimated by the lock detector metric shown in Fig. 14. The metric was defined by the following expression [2]:

$$M = \frac{IP \cdot IP - QP \cdot QP}{IP \cdot IP + QP \cdot QP} \cdot 100$$

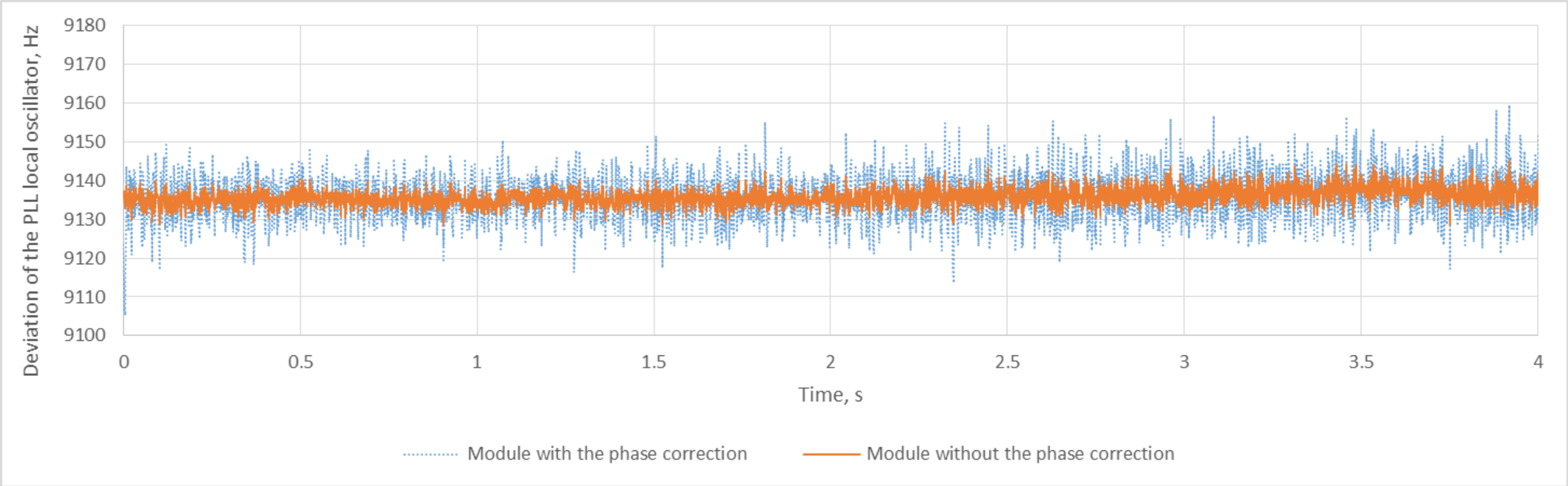

**Fig. 13.** Frequency of the local carrier oscillator

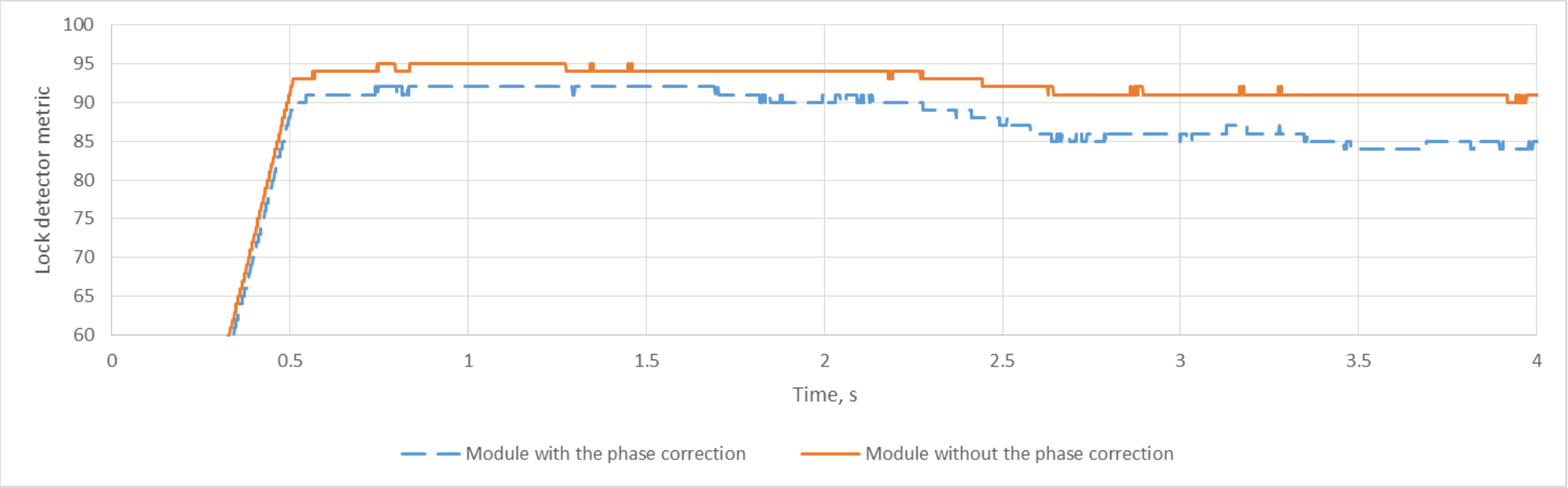

**Fig. 14.** Lock detector metric $C/N_0 \approx 38\ dbHz$

In this experiment, the lock detector was used to record the fact of tracking acquisition. The tracking module was considered in the lock state when the threshold value of 60 was reached. Since the tracking module strives to minimize the QP value, the state of ideal synchronization is achieved when QP = 0.

It should be noted that the negative effect of the metric decrease due to increased fluctuations might not be critical in practice, since at low SNR levels, integration over longer intervals is most often used. More important in this situation is the reliable tracking lock in the cases with a significant initial frequency deviation.

When increasing the integration time, the phase correction method allows the use of a narrower filter bandwidth. The issues of selecting the filter bandwidth in terms of improving tracking accuracy and stability are discussed in detail, for example, in [2, 3], and are not discussed here; however, phase correction allows improving the tracking process by using filters with a narrower bandwidth than it is possible in a module without phase correction.

Thus, in Fig. 15.1–15.4, the changes in the carrier frequency in the tracking modules are shown; the module with phase correction has a filter noise bandwidth of 5 Hz when integrating over 1 ms and 1 Hz when integrating over 20 ms (5/1), and the module without phase correction has bandwidths of 25 Hz and 5 Hz, respectively (25/5). The switching of the integration time occurs at approximately 5.7 s, which is visible in Fig. 15.1, where the graph of the local oscillator frequency for the entire interval is shown, and the effect of a relatively short-term interference at the end of the interval is also visible. Fig. 15.2 and 15.3 show the effect of the interference and the moment of integration time switching on an enlarged scale.

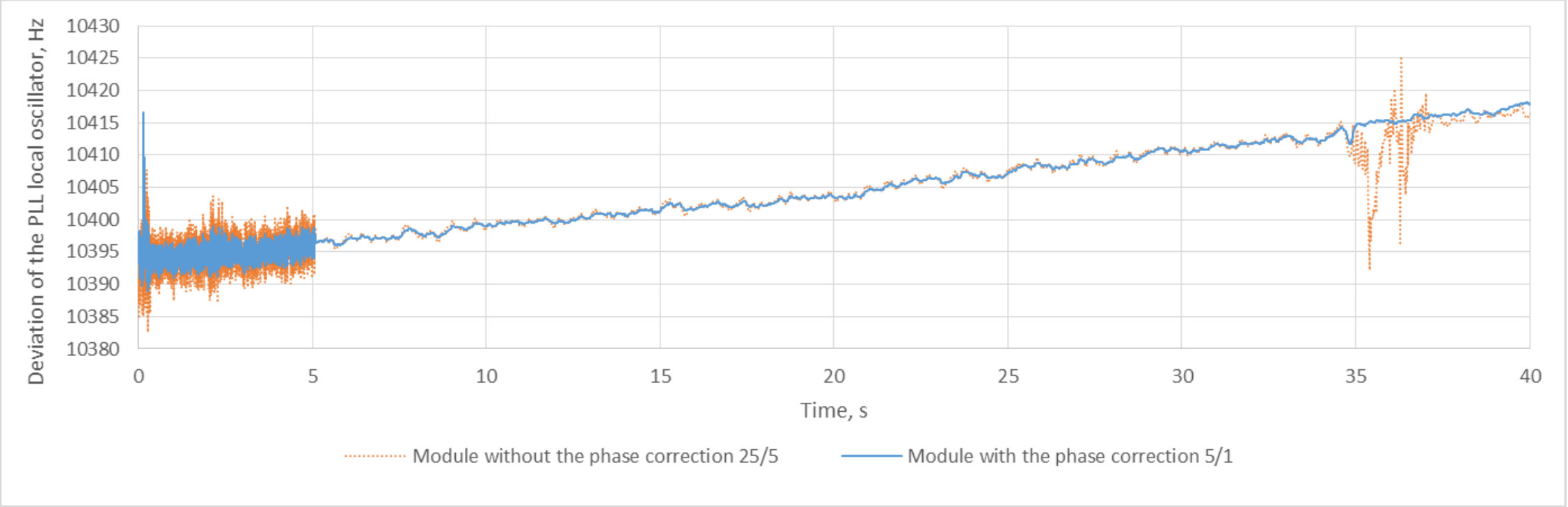

**15.1.** The dynamic behaviour of the tracking process

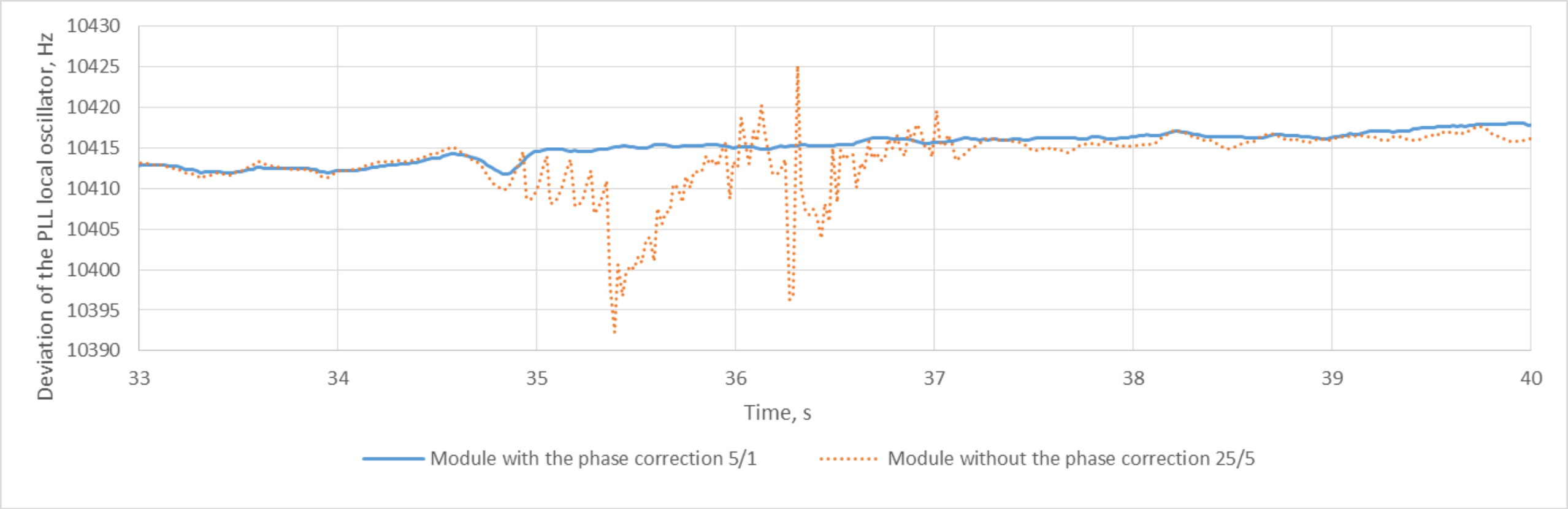

**15.2.** Short time interference

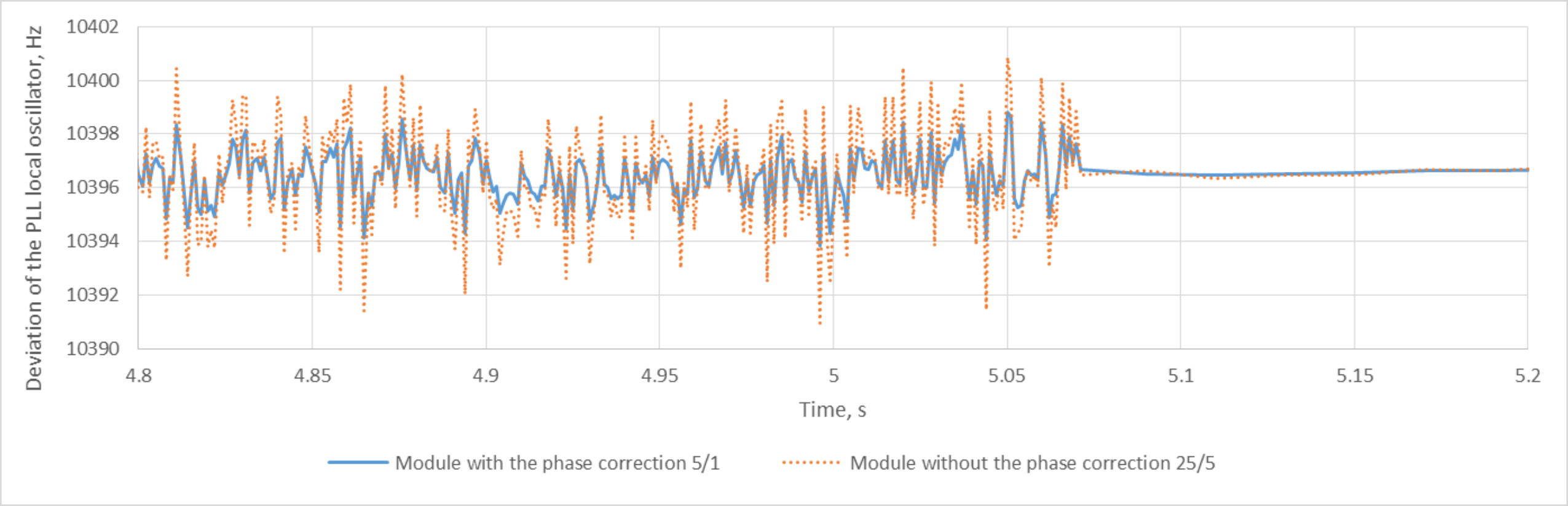

**15.3.** Switching from 1 ms mode to 20 ms mode

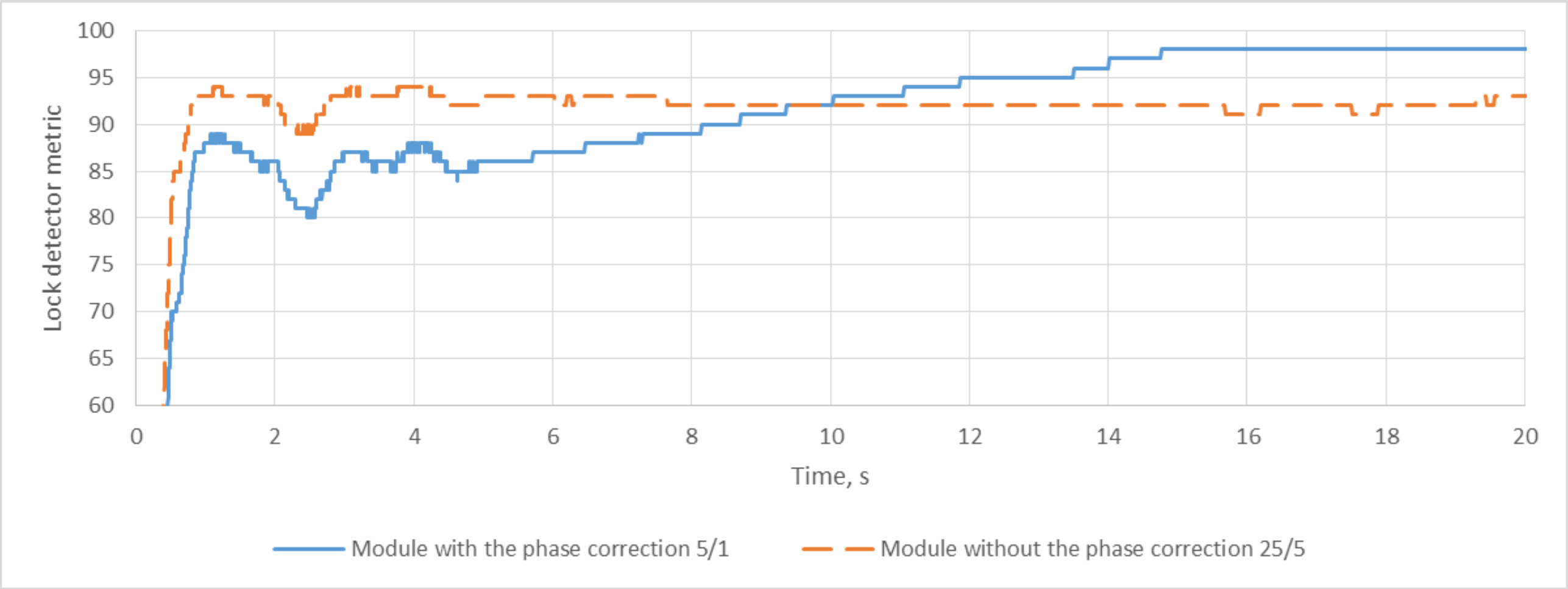

**15.4.** Metric of the lock detector (interval after 20 s is omitted)

**Fig. 15.** Estimation of the phase correction influence on the tracking process

From the graphs in Fig. 15, it can be concluded that the phase correction method has a positive effect on the tracking process with an increased integration interval, which is achieved by narrowing the filter noise bandwidth while maintaining stable tracking.

Fig. 16.1–16.3 show the results of comparing the operation of tracking modules with phase correction and filters with different bandwidths. In this experiment, the same satellite was simultaneously tracked by eight tracking modules with identical settings, except for the PLL filter bandwidth. The $C/N_0$ value was approximately 37-40 dBHz. Integration started with an interval of 1 ms, then switched to the 20 ms mode at approximately 5 s.

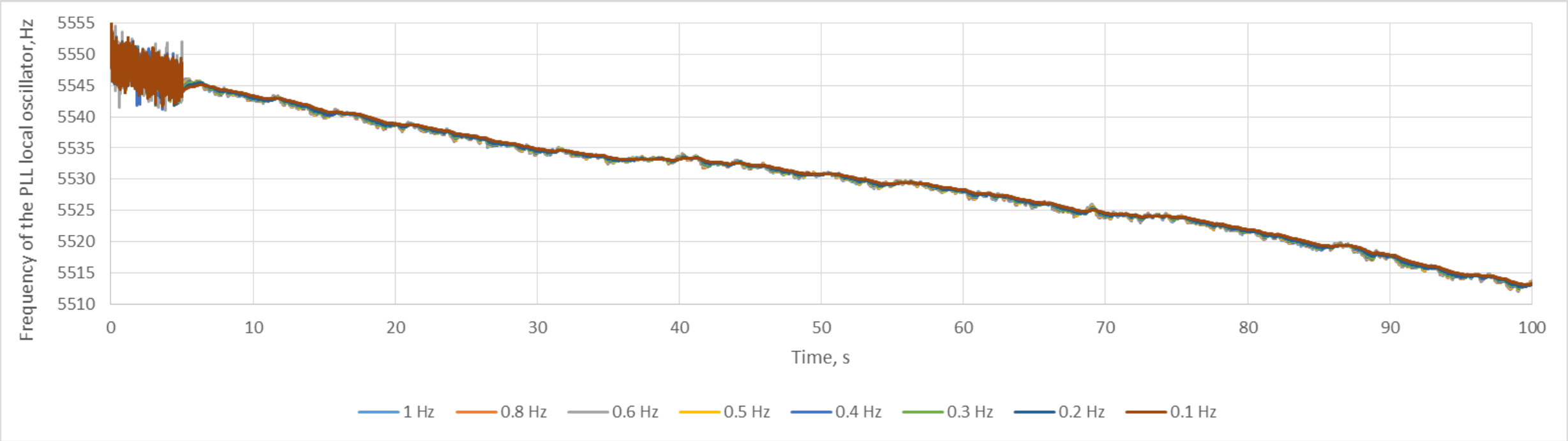

**16.1.** Frequency of the local carrier oscillator

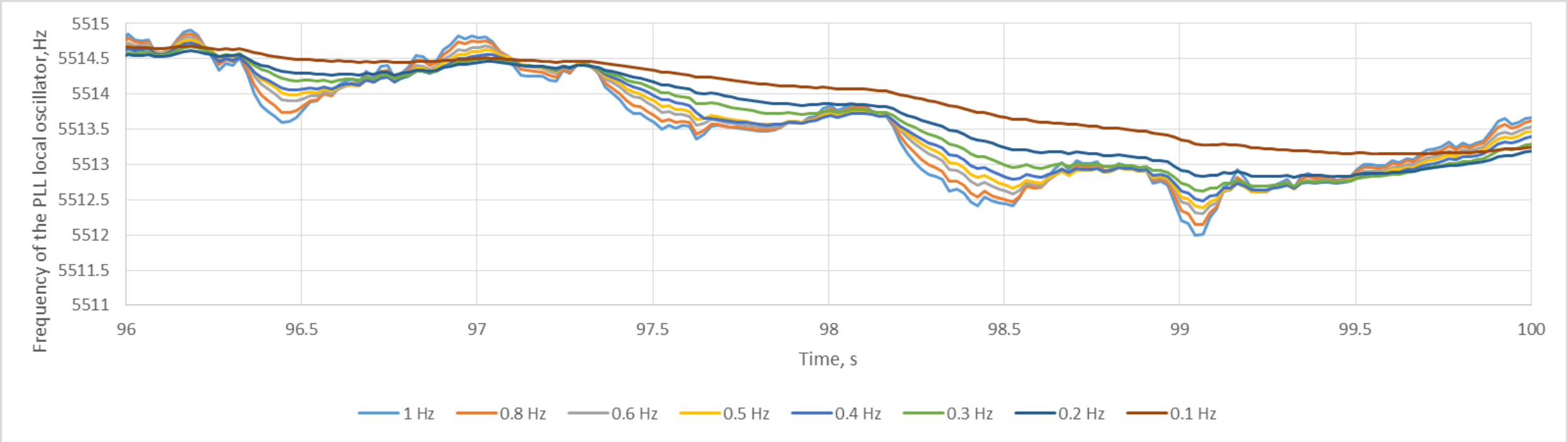

**16.2.** The magnified fragment of the graph after 96s (integration interval 20 ms)

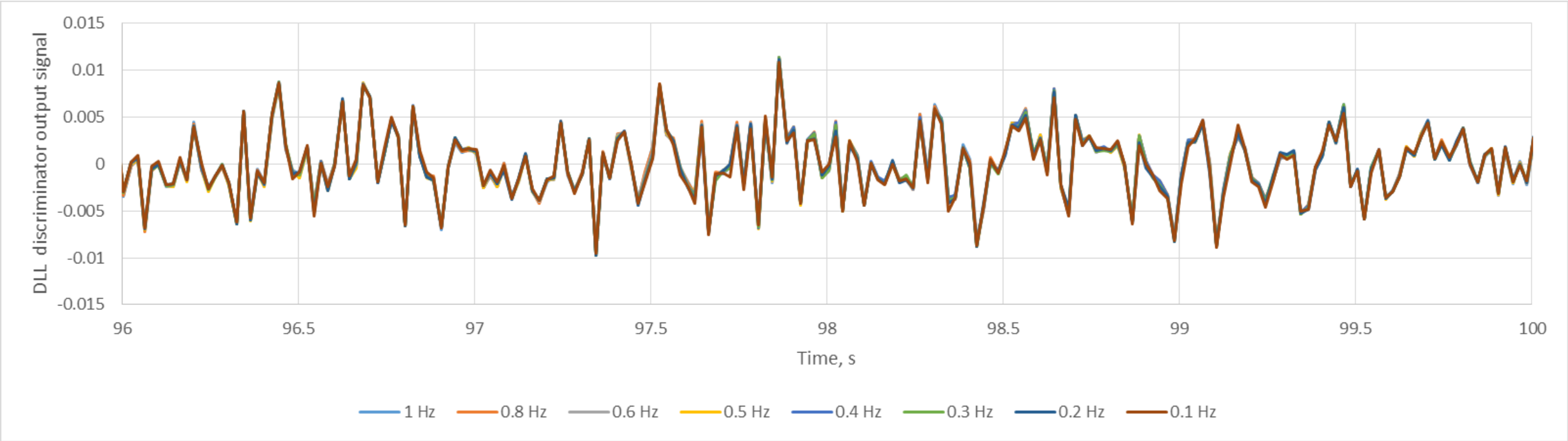

**16.3.** DLL discriminator output (integration interval 20 ms)

**Fig. 16.** For the choice of the noise bandwidth of the PLL filter in case of phase correction

The output signal of the discriminator in the DLL loop demonstrates the magnitude of the fluctuations of the local ranging code, which is subsequently converted into pseudoranges and thus characterizes the tracking error. The value of 0.01 corresponds to approximately 3 m. It can be seen that the tracking errors practically do not depend on the choice of the noise bandwidth in the range of 0.1–1 Hz. Further narrowing leads to an increase in the PLL inertia and can increase the tracking error, especially in the case of moving objects.

To evaluate the influence of the coefficient $a_2$ a testing set was implemented in which the same signal was fed to two tracking modules differing in the value of this coefficient. With a signal power of 35 dB-Hz and above, as well as with integration over a 20 ms interval, the difference in the tracking error value, both in the carrier loop and in the DLL, is practically absent; i.e., it can be concluded that it is possible to simplify the filter while simultaneously applying phase correction. It should be noted that with a weak signal, the module with a non-zero value of $a_2$ in some cases demonstrated higher stability; however, this was of no practical significance in our conditions, since signals with this $C/N_0$ did not allow ensuring stable synchronization.

In some cases, GNSS receivers use phase measurements to smooth pseudoranges or for other purposes; therefore, it is important to ensure the stable operation of the carrier phase meters. Fig. 17 shows a graph of the phase difference between modules configured as described above (eight modules with various filter bandwidths). The phase difference is reduced to the reference channel with a noise bandwidth of 1 Hz in accordance with the expression $\Delta\Psi_k = \Psi_k - \Psi_1$, where $\Psi_k$ is the phase of the signal with the bandwidth indicated in figure 17, and $\Psi_1$ is the phase of the channel with a bandwidth of 1 Hz.

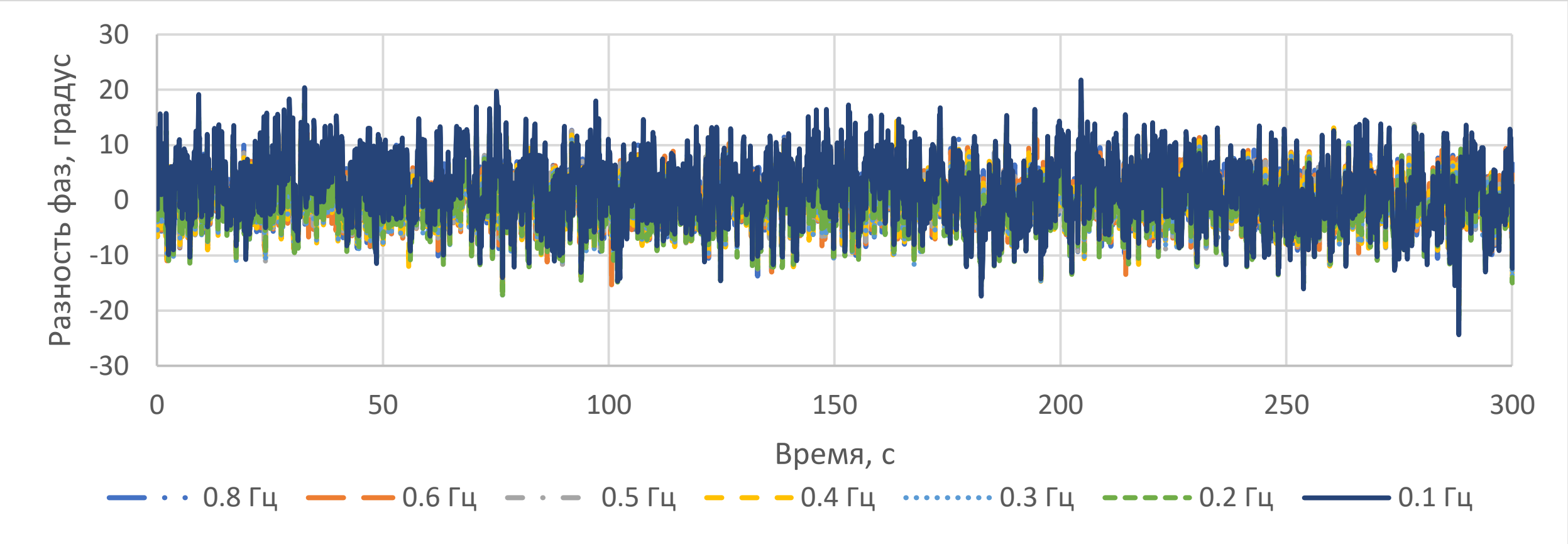

**Fig. 17.** Phase error in the PLL with various filter bandwidths (integration interval 20 ms)

Obviously, when using the proposed phase correction method, no disruptions in the operation of the phase meter are observed, and the influence of the filter noise bandwidth is insignificant. Nevertheless, choosing a filter bandwidth too narrow can lead to an additional error associated with an increase in the filter inertia.

## Conclusion

The results of simulation and experimental studies show that the proposed method of phase correction in the DDS-based local carrier oscillator positively affects the operation of the tracking module, especially when using integration intervals longer than 1 ms. The main effect is achieved by narrowing the filter bandwidth in the PLL loop and reducing the probability of phase difference accumulation, which can lead to additional errors. The paper considers the implementation of a modified second-order filter; however, there are no obstacles to extending the proposed method to the tracking modules with the third-order filters using appropriate correction of the filter parameters. A disadvantage of the method is the use of unfiltered PLL discriminator signals in the tracking loop, which, however, does not diminish its value when using extended integration intervals. In this situation it is reasonable to discuss the opportunity of using the nonlinear Kalman filtering [14] or similar solutions intended for the improvement of tracking performance in case of weak signals.

The use of the phase correction method allows significantly narrowing the noise bandwidth without losing stability. In particular, experimental studies have shown that with an integration interval of 1-5 ms, a bandwidth of about 5 Hz is sufficient; when integrating over 20 ms, the PLL loop bandwidth can be narrowed to about 0.5-1 Hz, and the DLL loop bandwidth to about 0.05 Hz.